\documentclass{pasj00}
\usepackage{natbib}
\usepackage{fancyvrb,color}

\begin{document}
\SetRunningHead{K. Tadaki et al.}{Cosmic Star Formation Activity at $z=2.2$}

\title{Cosmic Star Formation Activity at $z=2.2$ Probed by H$\alpha $ Emission Line Galaxies}

\author{Ken-ichi \textsc{Tadaki}\altaffilmark{1}, Tadayuki \textsc{Kodama}\altaffilmark{2,3}, Yusei \textsc{Koyama}\altaffilmark{1}, Masao \textsc{Hayashi}\altaffilmark{3},\\%
Ichi \textsc{Tanaka}\altaffilmark{2}, and Chihiro \textsc{Tokoku}\altaffilmark{4}
}
\altaffiltext{1}{Department of Astronomy, Graduate School of Science, University of Tokyo, Tokyo 113-0033, Japan}
\email{tadaki.ken@nao.ac.jp}
\altaffiltext{2}{Subaru Telescope, National Astronomical Observatory of Japan, 650 North A'ohoku Place, Hilo, HI 96720, USA}
\altaffiltext{3}{Optical and Infrared Astronomy Division, National Astronomical Observatory of Japan, Mitaka, Tokyo 181-8588, Japan}
\altaffiltext{4}{Astronomical Institute, Tohoku University, Aramaki, Aoba-ku, Sendai, Miyagi 980-8578, Japan}


%


\maketitle

\begin{abstract}
We present a pilot narrow-band survey of H$\alpha$ emitters at $z=2.2$ in the Great Observatories Origins Deep Survey North (GOODS-N) field with MOIRCS instrument on the Subaru telescope. The survey reached a 3$\sigma $ limiting magnitude of 23.6 (NB209) which corresponds to a 3$\sigma $ limiting line flux of 2.5 $\times$ 10$^{-17}$ erg s$^{-1}$ cm$^{-2}$ over a 56 arcmnin$^2$ contiguous area (excluding a shallower area). From this survey, we have identified 11 H$\alpha$ emitters and one AGN at $z=2.2$ on the basis of narrow-band excesses and photometric redshifts. We obtained spectra for seven new objects among them, including one AGN, and an emission line above 3$\sigma $ is detected from all of them. We have estimated star formation rates (SFR) and stellar masses ($M_{\mathrm{star}}$) for individual galaxies. The average SFR and $M_{\mathrm{star}}$ is 27.8~M$_\odot$yr$^{-1}$ and 4.0$\times 10^{10}M_\odot$, respectivly. Their specific star formation rates are inversely correlated with their stellar masses. Fitting to a Schechter function yields the H$\alpha $ luminosity function with log$L$ = 42.82, log$\phi$ = $-$2.78 and $\alpha$ = $-$1.37. The average star formation rate density in the survey volume is estimated to be 0.31~M$_\odot$yr$^{-1}$Mpc$^{-3}$ according to the Kennicutt relation between H$\alpha $ luminosity and star formation rate. We compare our H$\alpha $ emitters at $z=2.2$ in GOODS-N with narrow-band line emitters in other field and clusters to see their time evolution and environmental dependence. We find that the star formation activity is reduced rapidly from $z$=2.5 to $z$=0.8 in the cluster environment, while it is only moderately changed in the field environment. This result suggests that the timescale of galaxy formation is different among different environments, and the star forming activities in high density regions eventually overtake those in lower density regions as a consequence of ``galaxy formation bias'' at high redshifts.
\end{abstract}

\section{Introduction}
Since recent observations in optical and near-infrared wavelength indicate that the volume-averaged star formation rate (SFR) increases from $z = 0$ to $z \sim 1$ and plateaus at $z \sim  2$ \citep{2004ApJ...615..209H,2006ApJ...651..142H}, it is likely that a large fraction of stars in galaxies at present-day formed at $z>1$. The AGN activity and the redshift distribution of submm galaxies (SMG) also peak at this epoch \citep{2005ApJ...622..772C}. Therefore the redshift range of $z$=2--3 is the epoch when galaxies have the most intensive evolution. It is absolutely imperative to build a statistical sample of star forming galaxies at $z$=2--3 in order to understand the cosmic star formation history and the early evolution of galaxies. A line-emitter survey with a narrow-band filter can provide a large sample of active galaxies efficiently in a limited range of redshift in contrast to color-color selections such as $UG\mathcal{R}$ \citep{2004ApJ...604..534S} or $BzK$ \citep{2004ApJ...617..746D}. Since red galaxies represent characteristic colors on the color-magnitude diagram as referred to as the ``red sequence'', we can relatively easily select such red galaxies located at a specific redshift \citep{1998A&A...334...99K}. The blue star-forming galaxies, however, are much more difficult to be identified at $z\sim2$ using a color-selection or a photometric redshift technique because their spectra are relatively flat and featureless in the optical-NIR regime. Moreover, in a narrow-band survey which captures emission lines from galaxies directly, we can sample star forming galaxies completely above a certain limiting flux and an equivalent width limit unless the line is attenuated by dust or stellar absorption. We are not biased by colors of galaxies, either.  For these reasons, the narrow-band emitter survey is a very efficient and effective method for initially making a sample of star forming galaxies and obtaining their photometric properties at a particular redshift. We can then study them more in detail by spectroscopic follow-up in the near-infrared and also by radio observations targeting their molecular lines.

One of the most important quantities characterizing a galaxy is SFR. There are many different SFR indicators, such as UV continuum, nebular emission lines (H$\alpha $ and [OII]) and mid-infrared, but the SFRs estimated by various measurements do not always provide consistent results due to selection biases and different amounts of dust extinction effects, and so on. For UV continuum radiated by hot OB type stars, we must correct for absorption by the surrounding dust. This effect is often corrected for by using the UV slope but this process may lead to a large uncertainty in the estimated SFRs.  On the other hand, mid-infrared emission is not affected by dust ``extinction'' since it is dust ``emission''.  However, we can observe only the galaxies that have relatively high SFRs so as to be detected by mid-infrared observations which are less sensitive to optical or near-infrared observations. Also, to derive SFRs from the mid-infrared luminosity one has to assume the dust temperature which is somewhat uncertain. The H$\alpha $ line, a hydrogen's Balmer series line emitted from ionized gas (i.e., HII region) around hot young stars, is one of the best SFR indicators. It has many great advantages; being less affected by dust extinction, providing a survey with high sensitivity, and having been well calibrated in the local Universe. For the rest-frame UV-selected galaxies, \cite{2006ApJ...647..128E} obtained 114 H$\alpha $ spectra of star forming galaxies at $z \sim 2$ and indicate that H$\alpha $ emission is attenuated by a typical factor of 1.7, which is about 1/3 compared to the UV attenuation. 

The narrow-band survey of H$\alpha $ line enables us to make a relatively-unbiased large sample of star forming galaxies at the same redshift as well as to obtain accurate estimates of SFR, unlike the rest-frame UV surveys which tend to miss a significant fraction of star-forming activities due to dust obscuration. However, the H$\alpha $ line is red-shifted into the near-infrared regime at $z > 0.5$. Because the near-infrared observation is severely affected by OH sky emission lines, only a fraction of redshift range can be surveyed with high sensitivities. At $z\sim $2.2, H$\alpha $ line falls just in between the OH line forest ($\lambda \sim 2.1 \mu $m). There are some pioneering previous works that searched for H$\alpha $ emitters at $z\sim $2 with narrow-band imaging \citep{1994AJ....107....1T,1995MNRAS.273..513B,1998ApJ...506..519T} or with slitless spectroscopy \citep{1999ApJ...519L..47Y,2010ApJ...723..104A}. In the past decade, some large H$\alpha $ surveys have been carried out at $z \sim 2.2$. \cite{2000A&A...362....9M} searched for H$\alpha $ emitting galaxies with ESO NTT telescope. The survey reached a limiting line flux of 5$\times 10^{-17}$ erg s$^{-1}$ cm$^{-2}$ and covered a 100~arcmin$^2$ area. The observation with the narrow-band filter yielded 10 candidates and 6 of them have been confirmed spectroscopically later. Then, \cite{2008MNRAS.388.1473G} conducted a narrow-band survey with a line flux limit of $\sim 10^{-16}$ erg s$^{-1}$ cm$^{-2}$ over a large area of 0.6 deg$^2$ with a 3.8m telescope (UKIRT) and identified 55 H$\alpha $ emitters. Although their survey provided a sample large enough to evaluate the H$\alpha $ luminosity function (LF) at the bright end, they pointed out that the faint end of H$\alpha $ LF is important in order to estimate the global star formation rate, and hence a deeper survey is required for this purpose. To date, the deepest survey of H$\alpha $ emitters at this high redshift is that of \cite{2010A&A...509L...5H}. They carried out a narrow-band survey with HAWK-I on ESO-VLT, and reached to a 3$\sigma $ flux limit of 4.1 $\times 10^{-18}$ erg s$^{-1}$ cm$^{-2}$ over 56~arcmin$^2$ area. Their survey suggested that the faint-end slope of the H$\alpha $ LF is steeper than the value obtained in the local universe. However, large scale structures at $z\sim 2$ may stretch the intrinsic slope of the LF and it is too early to conclude on the faint end of H$\alpha $ LF by only the data in a single field. The result is very susceptible to the effect of cosmic variance.

Although the recent development of wide-field instruments in the near-infrared regime has enabled us to study distant galaxies at $z\sim 2$ systematically, there is an on-going debate as to how the ancient galaxies evolve into the ones at the present-day universe. This is attributed to the fact that the galaxy evolution is connected not only to time, but also to environment and mass \citep[e.g.][]{2007A&A...468...33E,2010A&A...518A..18T}. It is thought that there are two types of causes for the environmental dependence, inherent origin and acquired effects. In the former case, initial density fluctuation is originally a bit large in high density regions such as cluster cores, and galaxy formation take place earlier there compared to lower density regions. In the later case, some external factors, such as gas stripping by ram pressure and mergers between galaxies, influence galaxy properties during the course of their build-up. The spatial distribution of star forming galaxies helps us to tell whether the inherent origin or the acquired effects dominant the environmental dependence. In galaxy cluster RXJ1716 at $z=0.8$, H$\alpha $ emitters or mid-infrared sources, which are both tracers of active star formation, are preferentially found in the surrounding region of the cluster core \citep{2008MNRAS.391.1758K,2010MNRAS.403.1611K}. In contrast, star-forming galaxies in XCS2215 at $z=1.5$ are located in the cluster core as often as in the surrounding environments \citep{2010MNRAS.402.1980H,2010ApJ...719L.126T}. It seems that star formation activity is initially enhanced in massive galaxies in high density regions and is propagated to lower mass galaxies and to lower density regions with time. If the relative velocity between galaxies is too large, these galaxies would pass through each other without merging together. Therefore, at $z=0.8$, it is thought that mergers in mid-density regions would be more important than the inherent effect. However, at a higher redshift $z=1.5$, it seems that the inherent effect would surpass the acquired effects. To further look into and confirm this interesting result based on the two clusters, and to extend it further back in time, it is required to construct a larger sample of star forming galaxies over the critical era of $z=$1--3, that covers various environments from galaxy cluster core to blank field region in order to compare the properties between different environments.
 
In this paper, we present a survey of H$\alpha $ emitters at $z=2.2$ in GOODS-N field in order to study star formation activities in the general field or low density regions. This paper is structured as follows. In \S~2, we describe our survey design, observations, and the data. In \S~3, our target selection of H$\alpha $ emitters and the spectroscopic follow-up observations are described. We show our results in \S~4 and \S~5, and discuss the star formation activities of galaxies in various environments and redshifts in \S~6. Finally, we summarize our study in \S~7. We assume the cosmological parameters of H$_0$=70 km s$^{-1}$ Mpc$^{-1}$, $\Omega_M=0.3$, and $\Omega_\Lambda=0.7$, and adopt AB magnitudes throughout this paper.

\section{Data}
\subsection{Triple narrow-band filter system for $z=2.2$}
\begin{figure}
\begin{center}
 \FigureFile(80mm,60mm){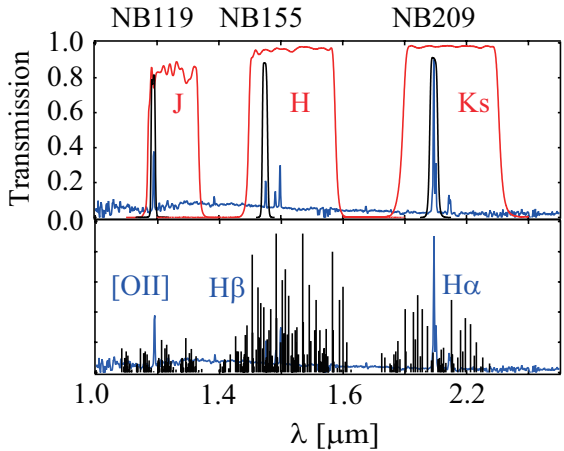}
\caption{The transmission curves of our triple narrow-band filter system (black) and broad-band filters (red). Lower panel shows the wavelength of OH sky emission lines \citep{2000A&A...354.1134R}. Blue solid line shows the template spectrum of a typical star-forming galaxy at $z=2.2$ \citep{1996ApJ...467...38K}.}
\label{fig1}
\end{center}
\end{figure}

Taking the advantage that H$\alpha $ line at $z=2.19$ is free from any strong OH sky emission line, we have made three sets of custom-made narrow-band filters, namely, NB119 ($\lambda _\mathrm{c}=1.19\mu $m, FWHM=0.014$\mu $m), NB155 ($\lambda _\mathrm{c}=1.55\mu $m, FWHM=0.017$\mu $m) and NB209 ($\lambda _\mathrm{c}=2.09\mu $m, FWHM=0.027$\mu $m). Figure 1 shows the filter response functions of these narrow-band filters together with those of broad-band filters $J$, $H$ and $K_\mathrm{s}$.  NB119, NB155 and NB209 filters capture [OII], H$\beta $ and H$\alpha $ emission lines, respectively, from the same galaxies at $z=2.19\pm 0.02$. If emission lines are detected in more than a single filter from a galaxy, we can confirm its redshift without spectroscopy. Even if only a single line is detected, all we need to do is to discriminate among [OII]/[OIII]/H$\alpha $ by using photometric redshifts. The combination of NB209/NB155 filters tries to measure Balmer decrement (H$\alpha $/H$\beta$) in order to accurately correct for dust extinction for the brightest emitters. In this paper, however, we will concentrate on the studies using the NB209 filter only, since we have not obtained deep enough data with NB155 yet.

\subsection{Imaging data}
The imaging survey has been carried out with MOIRCS \citep{2008PASJ...60.1347S} on the Subaru telescope \citep{2004PASJ...56..381I} with the NB209 filter in GOODS-N field. MOIRCS is equipped with two Hawaii-2 detectors (2048$\times $2048). Unfortunately, however, because one of the two MOIRCS chips was unavailable at the time of the observation run due to a trouble, one pointing could cover only a half of the original FoV that is 4\timeform{'} $\times $ 3.5\timeform{'}. We spent five pointings in total in April 2008 and the total area of the survey amounts to 70 arcmin$^2$, corresponding to about 9300 Mpc$^3$ in co-moving volume for the narrow-band survey at $z=2.19\pm0.02$. Table 1 shows the observed fields, exposure times, point spread functions (PSFs) and limiting magnitudes of NB209 and $K_\mathrm{s}$-band. Data reduction was conducted by using MCSRED (I.\ Tanaka, et al., in preparation). Selfsky flat image was used to correct for variabilities in sensitivity from pixel to pixel. Before combining all the images, PSFs were matched to 0.8\timeform{"} in GT4 and to 0.7\timeform{"} in other fields by gaussian smoothing. We calibrated the flux by observing a spectroscopic standard star, G191-B2B, and measured the photometric zero-point of our NB209 image (23.9). Limiting magnitude was estimated by using ``limitmag", which is a task of SDFRED \citep{2002AJ....123...66Y,2004ApJ...611..660O}. Note that the narrow-band image in GT4 is shallower than the other fields.
To select line emitters, and to measure their line fluxes, the continuum levels should be measured as well. As shown in Fig.\ 1, the NB209 filter is in the middle of the $K_\mathrm{s}$ filter and $K_\mathrm{s}$ magnitudes can be used to measure the continuum levels.
For $K_\mathrm{s}$ band data, we used the data-set of the MOIRCS Deep Survey (MODS; \citealt{2009ApJ...702.1393K}), which covers 70$\%$ of the GOODS-N field. PSF of the $K_\mathrm{s}$ band data was matched to the NB209 data. Also, we used the MODS photometric catalog (Kajisawa et al., in prep), which contains optical ($UBVi'z'$), near-infrared ($JHK_\mathrm{s}$), mid-infrared (IRAC 3.6, 4.5, 5.8, 8$\mu $m and MIPS 24$\mu $m) and X-ray data. The photometric redshifts and the stellar masses, based on GALAXEV \citep{2003MNRAS.344.1000B}, are also obtained from this catalog.

\begin{table*}
\begin{center}
\caption{Summary of the observations.}
\begin{tabular}{@{}llllllll@{}}
\hline
Field &$\alpha $&$\delta$&t$_{\mathrm{exp}}$& PSF&NB209&K&3$\sigma $ limiting line flux\\
&[J2000]&[J2000]&[s]&[\timeform{"}]&[AB,3$\sigma$,1.5\timeform{"}]&[AB,3$\sigma$,1.5\timeform{"}]&[$10^{-17}$erg s$^{-1}$cm$^{-2}$ ]\\
\hline
GT1 &12 36 11&+62 11 43&11220&0.7&23.6&25.1&2.5\\
GT2 &12 36 36&+62 14 20&10440&0.7&23.8&25.8&2.0\\
GT2R&12 36 57&+62 11 46& 8160&0.7&23.6&25.7&2.5\\
GT3 &12 37 01&+62 17 07&10020&0.7&23.7&25.3&2.2\\
GT4 &12 37 23&+62 19 47& 5040&0.8&23.0&25.2&4.2\\
\hline
\end{tabular}
\end{center}
\end{table*}

\begin{figure}
\begin{center}
 \FigureFile(80mm,50mm){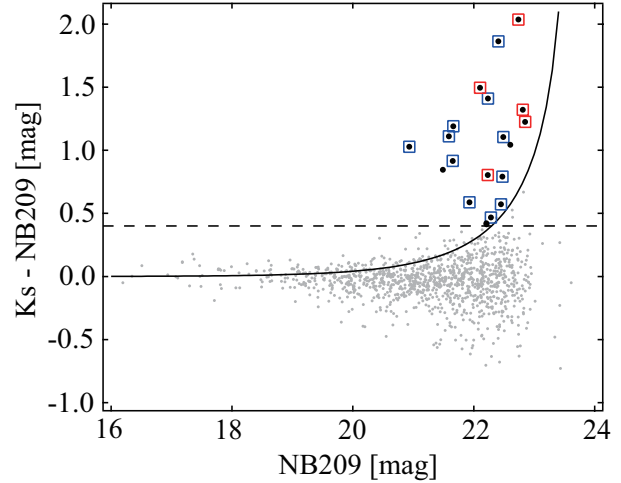}
\caption{The K - NB209 color-magnitude diagram to select H$\alpha $ emitter candidates at $z=2.2$. Grey dots show all the objects detected in both $K_\mathrm{s}$ and NB209 images and black dots show the NB209 emitters, which satisfy the criteria in section 3.1. H$\alpha $ emitters and [OIII] emitters are shown by blue and red squares, respectively. The solid curve and the dashed line show 3$\sigma $ photometric error and narrow-band excess of 0.4 magnitude in $K_\mathrm{s}$ - NB209, respectively.}
\label{fig2}
\end{center}
\end{figure}

\section{Target selection}

\subsection{Narrow-band emitters}

\begin{figure}
\begin{center}
 \FigureFile(80mm,40mm){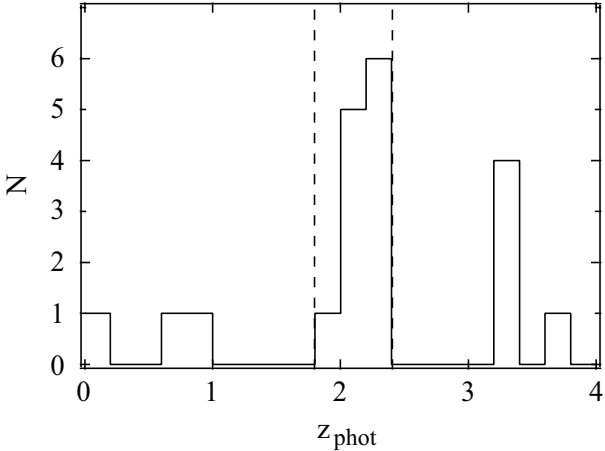}
\caption{The distribution of photometric redshifts for the NB209 emitters. The vertical dashed lines show our criteria to select H$\alpha $ emitters ($1.8<z_{\mathrm{phot}}<2.4$).}
\label{fig3}
\end{center}
\end{figure}

\begin{table*}
\begin{center}
\caption{Summary of our 12 H$\alpha $ emitters.}
\begin{tabular}{@{}lllllllll@{}}
\hline
ID	&		$\alpha $		&		$\delta$		&	$K_\mathrm{s}$	&	$f_{\mathrm{line}}$	&SFR&	$z_{\mathrm{spec}}$&	comment	\\
	&		[J2000]		&		[J2000]		&		&	[10$^{-17}$ erg s$^{-1}$ cm $^{-2}$ ]	&[M$_\odot \mathrm{yr}^{-1}$]&&		\\
\hline
MODS-91	&	189.0141 	&	62.1760 	&	23.07 	$\pm$	0.16 	&	2.28	$\pm$	1.82	&	12.2 	$\pm$	9.7 	&		&		\\
MODS-487	&	188.9930 	&	62.1988 	&	23.19 	$\pm$	0.17 	&	3.57	$\pm$	2.06	&	18.8 	$\pm$	10.8 	&	2.176 	&		\\
MODS-3653	&	189.0906 	&	62.2480 	&	22.65 	$\pm$	0.13 	&	3.09	$\pm$	2.25	&	16.7 	$\pm$	12.2 	&	2.200 	&		\\
MODS-6422	&	189.2159 	&	62.2513 	&	22.34 	$\pm$	0.11 	&				&				&	2.193 	&	AGN	\\
MODS-7651	&	189.2582 	&	62.2639 	&	23.37 	$\pm$	0.18 	&	5.93	$\pm$	2.22	&	31.6 	$\pm$	11.8 	&	2.188 	&		\\
MODS-7773	&	189.2364 	&	62.2788 	&	22.53 	$\pm$	0.13 	&	5.43	$\pm$	2.39	&	29.0 	$\pm$	12.8 	&	2.190 	&		\\
MODS-7889	&	189.2204 	&	62.2907 	&	23.53 	$\pm$	0.20 	&	3.87	$\pm$	1.91	&	20.6 	$\pm$	10.2 	&	2.187 	&		\\
MODS-7899	&	189.2235 	&	62.2901 	&	21.94 	$\pm$	0.09 	&	11.2	$\pm$	3.32	&	59.8 	$\pm$	17.7 	&	2.186 	&		\\
MODS-8313	&	189.2152 	&	62.3071 	&	24.00 	$\pm$	0.25 	&	3.98	$\pm$	1.71	&	21.2 	$\pm$	9.1 	&		&		\\
MODS-8540	&	189.2709 	&	62.2908 	&	22.63 	$\pm$	0.13 	&	8.54	$\pm$	2.83	&	45.6 	$\pm$	15.1 	&		&		\\
MODS-9441	&	189.3259 	&	62.2815 	&	22.43 	$\pm$	0.12 	&	4.26	$\pm$	2.46	&	22.9 	$\pm$	13.2 	&	2.196 	&		\\
MODS-10522	&	189.3564 	&	62.3194 	&	22.95 	$\pm$	0.15 	&	5.17	$\pm$	2.16	&	27.7 	$\pm$	11.6 	&	2.192 	&		\\
\hline
\end{tabular}
\end{center}
\end{table*}

Emission line objects were selected on the basis of the excesses of NB209 fluxes over $K_\mathrm{s}$ fluxes with the following criteria and procedures: (1) Source detection was performed on the $K_\mathrm{s}$ band image using SExtractor \citep{1996A&AS..117..393B}.
Usually, a narrow-band image is used for a detection. However, since our $K_\mathrm{s}$ band image is much deeper than the NB209 image by 1.5-2 magnitudes, we use $K_\mathrm{s}$ image for a detection. It is therefore unlikely that we miss many high equivalent width emitters with low continua. The extraction criteria were at least 25 pixels with fluxes above 2$\sigma $ level, where 1$\sigma $ is the sky noise of the image. (2) Marginal objects, whose fluxes were less than 5$\sigma $ in NB209 or less than 3$\sigma $ in $K_\mathrm{s}$ band, or those located at the edge of the image, were rejected as fakes. (3) In order to avoid statistically scattered objects around zero in $K_\mathrm{s}$-NB209 color due to photometric errors, we selected only the objects whose excesses were larger than 3$\sigma $ or 0.4 magnitude, corresponds to about 40\AA\ in the equivalent width in the rest frame (Fig.\ 2). Finally, 21 emitter candidates were selected.

It is expected that the NB209 filter detects not only H$\alpha $ emitters at $z=2.2$ but other line emitters at different redshifts. For example, there may be contamination of [OIII] emitters at $z=3.2$ and [OII] emitters at $z=4.6$. 
In order to separate H$\alpha $ emitters from the others, we used the spectroscopic redshifts and the photometric redshifts, taken from the MODS catalog. Four out of 21 candidates had spectroscopic redshifts, i.e.\ $z=$0.840, 2.186, 2.200 and 3.187. Figure 3 shows the distribution of the photometric redshifts for the NB209 emitters. This clearly exhibits a bimodal distribution at $z=2.2$ and $z=3.3$. The criterion of $1.8<z_{\mathrm{phot}}<2.4$ was adopted to select H$\alpha $ emitters, and we identified 12 H$\alpha $ emitters.  All of them also satisfy the $BzK$ criterion \citep{2004ApJ...617..746D}. One of them is likely to be a AGN because a X-ray emission is detected from this object \citep{2003AJ....126..539A}. The coordinates and the properties of these final emitters are listed in Table 2.

\subsection{Spectroscopic confirmation}

For the seven new emitters out of 12, including the AGN candidate, we obtained near-infrared spectra with MOIRCS on multi-object spectroscopy (MOS) mode \citep{2006SPIE.6269E.148T}. Our aims are to confirm that the narrow-band emitters are the H$\alpha $ emitters located at $z=2.2$, and to accurately measure their redshifts. The medium-resolution grizm, R1300+K, was used to possibly discriminate between H$\alpha $ and [NII] lines. The slit width was 0.7\timeform{"} and its length was 10-12\timeform{"}. The on-source exposure time was 320 minutes. Data reduction was conducted with the following procedures. First of all, a Sky frame was created by combining all the object frames.  We define sky noise as the poisson noise of this frame at each wavelength. The Sky frame and the object frames were then divided by the dome flat for flat fielding. Cosmic-rays and bad pixels were removed. Distortion was corrected by using the task ``mcsgeocorr" of MCSRED. The wavelength calibration was performed by using OH lines in the Sky frame.  For object frames, the Sky pattern was roughly removed by subtracting the adjacent dithered image. The residual background emission was subtracted by using the task ``background'' of IRAF. To calibrate the telluric absorption and the difference in sensitivity at each wavelength, the object frames were divided by the spectrum of a spectroscopic standard star (A0-type), and multiplied by the model spectrum of a A0-type star \citep{1998PASP..110..863P}. Flux is calibrated by using a star in the 2-MASS catalog \citep{2006AJ....131.1163S}.

Figure 4 shows their spectra in the rest frame. An emission line, whose flux is above 3$\sigma $, is detected from all of them. For MODS-7773, we can also see a weak [NII]$\lambda 6584$ line, hence these objects must be H$\alpha$ emitters for sure. For the others, [NII] lines do not appear. However, if a NB209 emitter is a [OIII] emitter at $z$=3.2, the doublet lines of [OIII]$\lambda \lambda 4959,5007$ (in the rest frame) or $\lambda _{\mathrm{obs}}$=2.083, 2.103 (in the observed frame) should have been resolved, given the resolution of our observations. But we do not detect any. Moreover, since the ratio of the [OIII] doublet is 1/3 \citep{2003AJ....125..514A}, we can reject the possibility of [OIII] emitters at $z=3.2$ for MODS-7651, 9441 and 10522, whose lines are detected at more than 5$\sigma$, and the weaker line should have been visible.  If any of them is a [OII] line at $z=4.6$, SFR based on the [OII] flux, would be estimated to be about 200 M$_\odot $yr$^{-1}$. Despite of such very high SFRs, however, these objects are not detected by MIPS 24$\mu$m ($S_{24}<80\mu $Jy). Therefore, these objects are also likely to be H$\alpha $ emitters at $z=2.2$. MODS-6422 is thought to be an AGN due to the X-ray detection. The H$\alpha $ profile looks marginally extended and the FWHM of the velocity width is about 850km s$^{-1}$, which is more than twice wider than those of other H$\alpha $ emitters. It is thus confirmed that our criteria, satisfying the NB209 excess and $1.8<z_{\mathrm{phot}}<2.4$, is effective and robust in removing contamination and selecting genuine H$\alpha $ emitters at $z=2.2$. In order to make more accurate confirmation, however, one would need to detect [OII] lines, which redshift to 1.19$\mu$m at $z=2.2$, by another spectroscopic follow-up observation in the J-band.

\begin{figure*}
\begin{center}
\begin{minipage}{140mm}
 \FigureFile(140mm,100mm){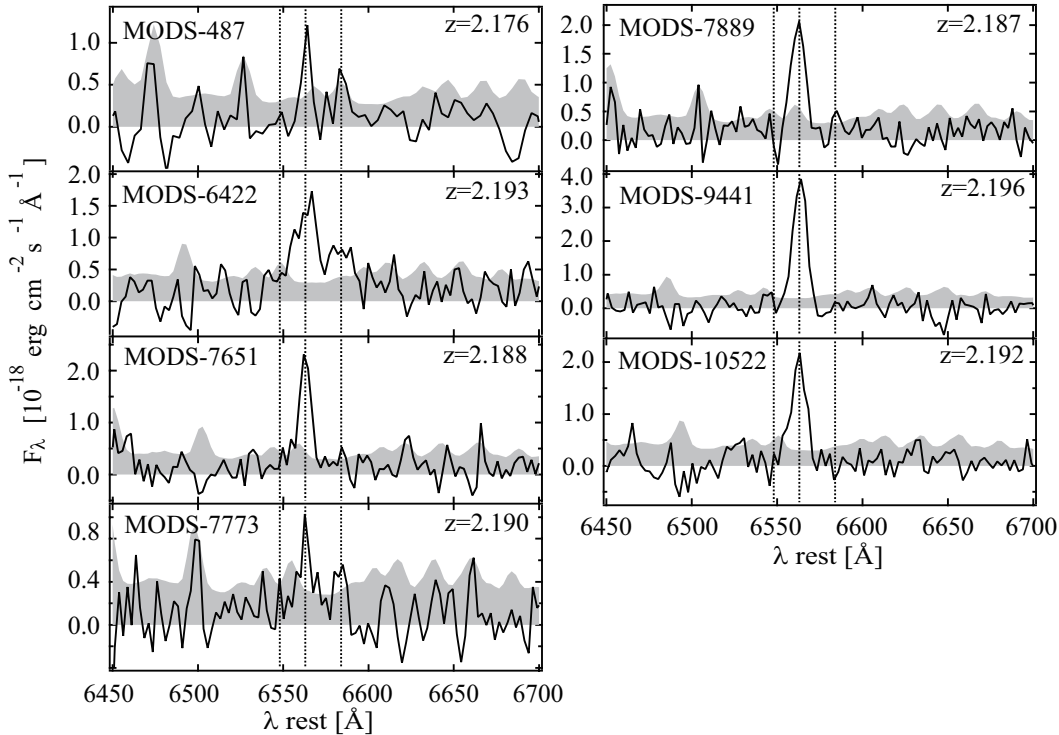}
\caption{H$\alpha $ spectra of the emission-line galaxies. These spectra have a binned dispersion of 8$\AA$ in the observed frame. For MODS-487 only, the bin size is 12$\AA$. The vertical dotted lines show the wavelengths of H$\alpha $ and [NII]$\lambda 6548, 6584$ lines. The gray regions indicate the 1$\sigma $ poisson noise.}
\label{fig4}
\end{minipage}
\end{center}
\end{figure*}

\section{Star formation rate and stellar mass}

We have constructed a sample of 11 H$\alpha $ emitting star-forming galaxies at $z=2.2$ in the GOODS-N field. We now estimate their SFRs and stellar masses. From the flux densities in NB209 and K$_\mathrm{s}$ band, i.e.\ $f_{\mathrm{NB}}$ and $f_{\mathrm{BB}}$, respectively, H$\alpha $ line fluxes can be calculated as follows,
\begin{equation}
F_\mathrm{line}=\frac{\Delta _{\mathrm{NB}} \Delta _{\mathrm{BB}}}{\Delta _{\mathrm{BB}} - \Delta _{\mathrm{NB}}}(f_{\mathrm{NB}}-f_{\mathrm{BB}}),
\end{equation}
where $\Delta $ indicates a wavelength width (FWHM) of a filter and $f$ denotes the flux density. We use $\Delta_{\rm NB}$=0.027[$\mu $m] and $\Delta_{\rm BB}$=0.31[$\mu $m]. We note that the [NII] line also contributes to a NB209 flux. We adopted L([NII])/L(H$\alpha $+[NII])=0.25 for all the objects \citep{2009MNRAS.398...75S}. Using the standard calibration of \cite{1998ARA&A..36..189K}; SFR[M$_\odot $yr$^{-1}$]=$7.9\times 10^{-42}$~L($\mathrm{H\alpha }$)[erg s$^{-1}$], we converted the H$\alpha $ luminosity to a SFR, assuming the Salpeter IMF \citep{1955ApJ...121..161S}. To correct for dust extinction, we derived $E(B-V)_{\mathrm{stellar}}$ from the SED fitting. In the local universe, it is known that emission lines from the ionized gas are more attenuated than stellar continua such as $E(B-V)_{\mathrm{stellar}}=0.44E(B-V)_{\mathrm{gas}}$ \citep{2000ApJ...533..682C}. On the other hand, at $z\sim2$, Erb et al. (2006) used the same extinction values for both stellar continua and the nebular emission lines, i.e.\ $E(B-V)_{\mathrm{stellar}}=E(B-V)_{\mathrm{gas}}$. This is because the above formula by Calzetti et al.\ (2000) significantly overestimates the H$\alpha $-based SFRs with respect to the UV-based SFRs.  The average strength of dust extinction for our H$\alpha $ emitters is estimated to be A(H$\alpha $)=0.7-1.6 if the Calzetti formula is used.  In this paper, we adopt the constant value of A(H$\alpha $)=1, given the uncertainty. As a future work, however, the amount of reddening should be ideally estimated from the Balmer lines ratio, such as H$\alpha $/H$\beta$. For emitters with spectroscopic redshifts, we also correct for the filter transmission depending on the location of the H$\alpha $ line on the response function of the narrow-band filter. For those without spectroscopic redshifts, we assumed the average redshift of $z=2.190$. The stellar masses were obtained from the MODS catalog \citep{2009ApJ...702.1393K} which were derived from the SED fitting. The average SFR and $M_{\mathrm{star}}$ of our H$\alpha $ emitters are 27.8 $M_\odot$ yr$^{-1}$ and 4.0$\times 10^{10}M_\odot$, respectively. 

The specific star formation rate (SSFR), defined as the SFR per unit stellar mass, shows how much the current star formation activity contributes to increase the stellar mass of a galaxy. For UV and near-infrared selected galaxies, it is reported that there is a good correlation between SSFR and stellar mass over a wide range in the evolutionary stage from young galaxies which begin to form stars, to passively evolving galaxies \citep{2006ApJ...647..128E, 2006ApJ...653.1004R}. SSFR is high in a less massive galaxy and it decreases with increasing stellar mass. Figure 5 shows SSFR versus stellar mass for our H$\alpha $ emitters. Open circles show the UV-selected galaxies of \cite{2006ApJ...647..128E,2006ApJ...646..107E}. Because they adopted the Chabrier IMF \citep{2003PASP..115..763C}, we convert their data points to those for the Salpeter IMF by multiplying 1.8 to the SFRs. For our H$\alpha $ emitters, the SSFR negatively correlates roughly with the stellar mass as seen in the samples of Erb et al. (2006a,b). Our samples have systematically lower SSFRs than those of the UV-selected samples. Two selection effects would be contributing to this offset. One is that our survey is deeper and can detect galaxies with much weaker SFRs down to $\sim$ 12~M$_\odot $yr$^{-1}$. In other words, it may merely reflect the difference in the flux limits. The other is the large difference in the survey volumes. Because the survey area of Erb et al. (2006a,b) is very wide, 0.48~deg$^2$ \citep{2004ApJ...604..534S}, corresponding to $6.1\times 10^6$~Mpc$^3$ in comoving volume at $z=1.4-2.5$, their survey can efficiently detect relatively rare objects with strong SFRs (e.g., SFR $\sim$ 100~M$_\odot $yr$^{-1}$). Based on the luminosity function derived in \S~5, we can expect to detect only 2 high SFR galaxies ($>100$~M$_\odot $yr$^{-1}$) in our survey volume (9300~Mpc$^3$). Therefore, the non-detection can be simply due to a cosmic variance effect. The inverse of SSFR indicates the star formation time-scale to form all the stars in a galaxy. The average time-scale of the H$\alpha $ emitters is $\langle\tau \rangle=1.4$~Gyr, which is smaller than the age of the universe at $z$=2.2 (3~Gyrs). Therefore star formation history of our H$\alpha $ emitters does not necessarily require a star-burst and can be reproduced even by a constant star formation model.

\begin{figure}
\begin{center}
  \FigureFile(80mm,50mm){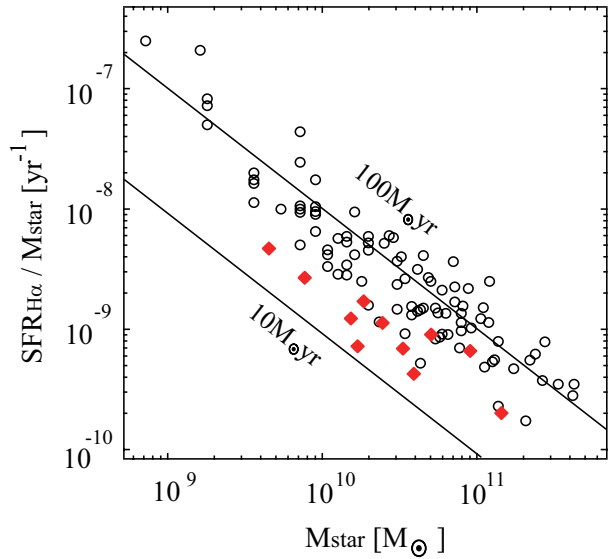}
\caption{The specific star formation rates (SSFR) versus stellar masses. The red diamonds and the open circles indicate our H$\alpha $ emitters and the UV-selected galaxies \citep{2006ApJ...647..128E,2006ApJ...646..107E}, respectively. We assumed the dust extinction of A(H$\alpha $)=1 [mag] and Salpeter IMF for both samples.}
\label{fig5}
\end{center}
\end{figure}

\section{Luminosity function and cosmic star formation rate density}

In order to estimate the volume-averaged star formation rate density, it is necessary to extrapolate the luminosity function to the unobserved faint star-forming galaxies. The H$\alpha $ LF is well represented by a Schechter function \citep{1976ApJ...203..297S}. In the local universe, the faint-end slope is measured to $\alpha $ = $-$1.3 \citep{1995ApJ...455L...1G}. However, recent studies indicate steeper faint-end slopes ($\alpha \sim $ -1.6) in the distant universe \citep{2010A&A...509L...5H}. Our deep survey can constrain the faint-end slope of H$\alpha $ luminosity function, independently. 
The volume density of galaxies is computed as
\begin{equation}
\Phi =\frac{1}{\Delta \mathrm{log} L}\sum \frac{1}{V_{\mathrm{max}}},
\end{equation}
within a luminosity bin of $\Delta $ log L = 0.4. $V_{\mathrm{max}}$ is the maximum co-moving volume probed by the survey. Because the filter response function does not have a perfect rectangle shape, $V_{\mathrm{max}}$ is larger for more luminous galaxies.
We have calculated the luminosity function excluding the GT4 field because the depth
of the narrow-band images in this filed is significantly shallower than the other fields.

\begin{figure}
\begin{center}
  \FigureFile(80mm,50mm){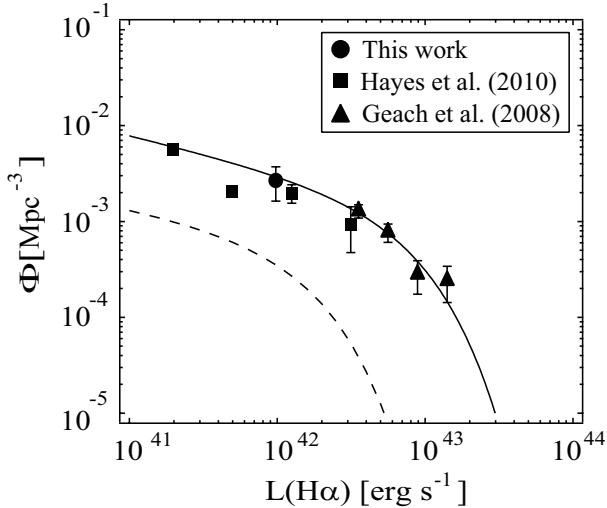}
\caption{The H$\alpha $ luminosity function at $z\sim 2.2$. The H$\alpha $ luminosity has not been corrected for dust extinction. The solid curve shows the best fit Schechter function with log$L$ = 42.82, log$\phi = -2.78$ and $\alpha = $-1.37. The dashed line shows the H$\alpha$ LF in the local universe with log$L$ = 42.15, log$\phi = -3.2$ and $\alpha = $-1.30 \citep{1995ApJ...455L...1G}. The faintest bin in this work includes seven H$\alpha $ emitters.}
\label{fig6}
\end{center}
\end{figure}

Figure 6 shows the H$\alpha $ luminosity function at $z\sim2.2$ obtained by this survey.
H$\alpha $ luminosity  has not been corrected for dust extinction. For the bright end, H$\alpha $ emitters of Geach et al. (2008) are used. Fitting a Schechter function to the data, without any correction for dust extinction, gives log$L = 42.82\pm 0.65$, log$\phi = -2.78\pm1.08$ and $\alpha = -1.37\pm0.44$. Our result does not support the steep slope at the faint end. This indicates that the cosmic variance has a considerably large impact on the result. Figure 7 shows the spatial distribution of H$\alpha $ emitters identified by this survey. Clearly, they are not homogeneously distributed across the field. This is not explained by the difference in the depth of the imaging data among different fields (Table 1). Most of our samples are concentrated in the GT3 field and there are only a few emitters in the other fields. We have calculated a LF in GT3 field only to investigate the influence of the cosmic variance. Fitting a Schechter function yields log$L$ = 43.36, log$\phi = -3.70$ and $\alpha = -2.08$, which now supports the steep slope at the faint end. This suggests that it is difficult to derive a representative number density of H$\alpha $ emitters at $z=2.2$ by a small survey ($<$100 arcmin$^2$). In contrast, the survey area ($\sim 0.6$~deg$^2$) of Geach et al. (2008) is wide enough to average the bright end of LF over the large-scale structures at $z=2.2$. It is thus also required to have an additional deep survey (L$_{\mathrm{H}\alpha } \sim 10^{42}$[erg s$^{-1}$]) over a much wider area to negate the effect of cosmic variance.

\begin{figure}
\begin{center}
  \FigureFile(80mm,50mm){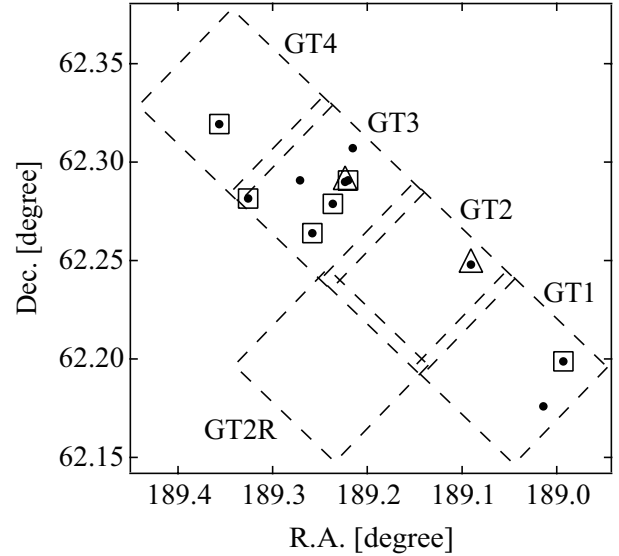}
\caption{The 2-D distribution of the H$\alpha $ emitters (filled circles). Triangles and squares show the spectroscopically confirmed objects at $z=2.2$ in Reddy et al.(2006) and in this work, respectively.}
\label{fig7}
\end{center}
\end{figure}

From the above derived H$\alpha $ LF, we can estimate the cosmic star formation rate density (SFRD). The H$\alpha $ LF is integrated over the range of $10^{37}<$L$_{\mathrm{H}\alpha }$[erg s $^{-1}$]$<10^{47}$ as in \cite{2004ApJ...615..209H}. The resultant volume-averaged SFRD is $\rho _{\mathrm{SFR}}=0.31$~M$_\odot$yr$^{-1}$Mpc$^{-3}$. If it is integrated only over the range of $10^{42}<$L$_{\mathrm{H}\alpha }$[erg s $^{-1}$]$<10^{47}$, the resultant SFRD is 0.21~M$_\odot$yr$^{-1}$Mpc$^{-3}$. If the faint end slope of the LF is steeper, the proportion of the extrapolation at the faint end would increase. The derived SFRD would be nearly doubled in the case of $\alpha =-1.6$. The evolution of SFRD has been studied with various indicators. We compare our result with those of previous studies with H$\alpha $ narrow-band imaging, and show the evolution of SFRD in Figure 8 \citep{2003ApJ...591..827P,2008AJ....135.1412D,2008PASJ...60.1219M,2007ApJ...657..738L,2008ApJS..175..128S,2003ApJ...586L.115F,2003A&A...402...65H,2008ApJ...677..169V,2009MNRAS.398...75S,2000A&A...362....9M,2008MNRAS.388.1473G,2010A&A...509L...5H}. The SFRD increases up to $z\sim 1$ but there is little change between $z=1$ and $z=2$. We note that this trend is also consistent with the result based on the {\it spectroscopic} sample at $z$=0.8, that is 0.1~M$_\odot$yr$^{-1}$Mpc$^{-3}$ \citep{2004MNRAS.354L...7D,2006MNRAS.370..331D}, although the direct comparison is difficult due to differences in sample selection and completeness. According to other star formation indicators, the SFRD does not increase but rather decreases at z$>3$ \citep{2006ApJ...651..142H}. Because H$\alpha $ shifts to $\lambda >2.6\mu $m at $z>3$, it is difficult to detect H$\alpha $ emission lines from distant galaxies with the ground-based telescope.
The space infrared telescopes such as JWST and SPICA \citep{2008SPIE.7010E..15N} will be very powerful for tracking the precise star formation histories further back in time at $z>3$.

\begin{figure}
\begin{center}
  \FigureFile(80mm,50mm){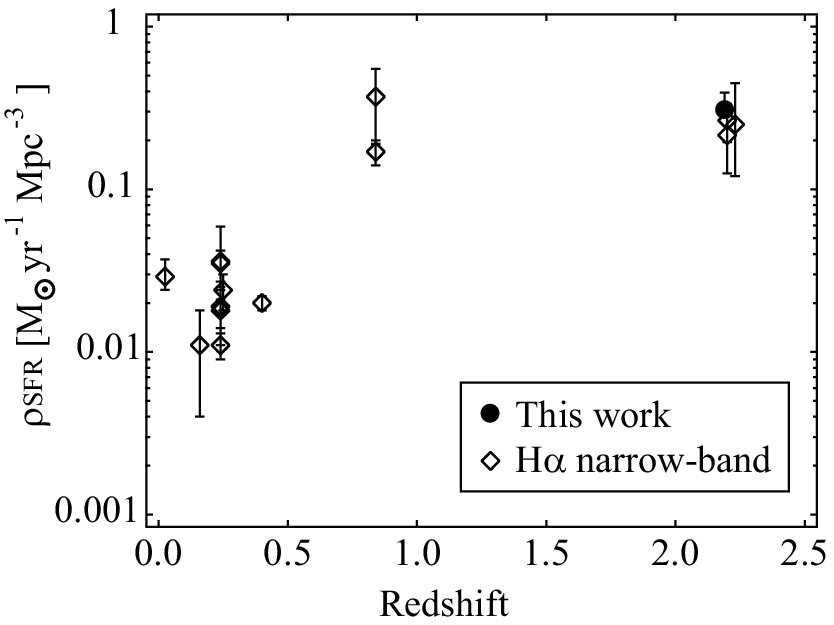}
\caption{The evolution of cosmic star formation rate density (SFRD). The filled circle and the open diamonds show SFRDs of this work and the previous studies of H$\alpha $ narrow-band imaging \citep{2003ApJ...591..827P,2008AJ....135.1412D,2008PASJ...60.1219M,2007ApJ...657..738L,2008ApJS..175..128S,2003ApJ...586L.115F,2003A&A...402...65H,2008ApJ...677..169V,2009MNRAS.398...75S,2000A&A...362....9M,2008MNRAS.388.1473G,2010A&A...509L...5H}.}
\label{fig8_1}
\end{center}
\end{figure}
\section{Star formation history as functions of redshift and environment}

As mentioned in \S~1, the evolution of galaxies depends strongly on the environment. Because field galaxies are antithetical to cluster galaxies in terms of environment, we can directly examine the environmental dependence of star formation activity by comparing the properties of emitters among different environments. In addition to the present general field survey, we have estimated SFRs and stellar masses for a large sample of star forming galaxies both in clusters and in the other general field at various redshifts (Field: [OII] emitters at $z=1.2$ in COSMOS, \citealt{2009ApJ...700..971I}; Clusters: H$\alpha $ emitters at $z=0.8$ in RXJ1716, \citealt{2010MNRAS.403.1611K}; [OII] emitters at $z=1.5$ in XCS2215, \citealt{2010MNRAS.402.1980H}). All the samples are selected on the basis of narrow-band nebular emission line surveys (H$\alpha $ or [OII]). We have limited the galaxy sample to those having SFRs above 12~M$_\odot$yr$^{-1}$ and stellar masses above $10^{10}$~M$_\odot$ to make fair comparisons.
We have estimated the stellar masses of the emitters with using the theoretical mass-to-luminosity ratios predicted by the population synthesis model \citep{1999MNRAS.302..152K} after it is scaled to the Salpeter IMF for consistency. At $z=2.2$,
\begin{equation}
\mathrm{log}(M_\mathrm{star}/10^{11})=-0.4(K^{tot}_\mathrm{s}-K^{11}_\mathrm{s})
\end{equation}
\begin{equation}
\Delta \mathrm{log}M=0.229[(z-K_\mathrm{s})-3.45].
\end{equation}
$z$ and $K_\mathrm{s}$ indicate $z$-band and $K_\mathrm{s}$-band magnitudes, respectively. $K^{11}_\mathrm{s}$ denotes the $K_\mathrm{s}$-band magnitude for a galaxy that has a stellar mass of $10^{11}$~M$_\odot$, and it is estimated to be 22.42 at $z=2.2$. The upper equation exhibits the mass--magnitude relation in the case of a passively evolving galaxy. The lower equation gives an amount of correction for stellar mass depending on the $z-K$ color, with respect to the mass given by the upper equation. The mass-to-light ratios of galaxies are different from redshift to redshift and among different colors, and are estimated by the model. The H$\alpha$/[OII] ratio is assumed to be log (H$\alpha$/[OII])$_{\mathrm{obs}}$=0.2 which is the observed average value in the local universe \citep{2006ApJ...642..775M}.
The ratio may actually depend on the luminosity (e.g.\ $B$-band in the rest-frame as in Moustakas et al. 2006), and adopting the fixed value may introduce some uncertainties. We need to examine the correlation between the H$\alpha$/[OII] ratio and the luminosity (or metallicity) for star-forming galaxies at high redshift with extensive near-infrared spectroscopy as a future work.
Figure 9 shows cumulative functions of SFR (left) and SSFR (right) of various samples. Blue lines and red lines correspond to field and cluster environments, respectively. We see an evolution of SFR function in the cluster environment in the sense that SFRs at higher-$z$ are systematically higher. However, SSFRs do not show a significant change over time. This indicates that the mass of galaxies which host star formation at $z=0.8$ is smaller than that of $z=1.5$. This is equivalent of the ``down-sizing'' effect \citep{ 1996AJ....112..839C,2004MNRAS.350.1005K, 2005ApJ...621..673T} where star-forming activity is shifted to low-mass galaxies with time.
Also, there is a clear difference between the two environments in the sense that 
SSFRs in the field environment are systematically higher than those in the clusters.

In order to investigate the star formation activities at higher redshift, we make a comparison with the properties of galaxies in distant proto-clusters (H$\alpha $ emitters at $z=2.2$ in PKS1138, Kurk et al. 2004; H$\alpha $ emitters at $z=2.5$ in 4C23.56, Tanaka et al. submitted). Note however that the limiting SFRs for 4C23.56 only ($\sim$ 25~M$_\odot $yr$^{-1}$) is significantly larger than our limit for the other fields/clusters (12~M$_\odot $yr$^{-1}$), and the emitter samples in 4C23.56 do not include galaxies with lower SFRs. In Figure 10, we present the evolution of star formation activities both in the fields and in the clusters. Vertical axis shows the ratio of total star formation rate to total stellar mass, which is equivalent to the averaged specific star formation rate. There is a difference in the slope of evolution between the two environments. These two lines seem to cross at $z\sim 2.5$. This may indicate that, at high redshifts, we may reach to the epoch when star formation activities become higher in higher density regions than the general field, as opposed to the local relation.

According to the Schmidt law \citep{1998ApJ...498..541K}, the SSFR can be related with the cold gas fraction, 
\begin{equation}
SSFR\propto \frac{\mu ^{1.4}}{1-\mu}.
\end{equation}
where $\mu \equiv M_{\mathrm{gas}}/( M_{\mathrm{gas}}+ M_{\mathrm{star}})$. In high density regions such as galaxy clusters, the gas fraction rapidly decreases from $\mu \sim 0.4$ at $z=2.5$ to $\mu \sim 0.1$ at $z=0.8$. Here we determined the proportional constant by using the observed values of $\langle \mathrm{SSFR} \rangle=3~\mathrm{Gyr}^{-1}$ and $\langle\mu \rangle=0.5$  for the sample of \cite{2006ApJ...644..792R}. It implies that these galaxies actively form stars and consume a massive amount of gas on a shorter time-scale due to the environmental effect. In this argument, we have assumed a closed box model and ignored any merger effect which would affect the time-scale and complicate the argument. However, our main motivation here is to simply and roughly discuss the global evolution in SSFR and gas fraction. Note also that SSFR and gas fraction would not change if two identical galaxies merge together. Therefore our discussion would not be significantly affected unless two galaxies at very different evolutionary stages merge together.

\begin{figure}
\begin{center}
  \FigureFile(80mm,50mm){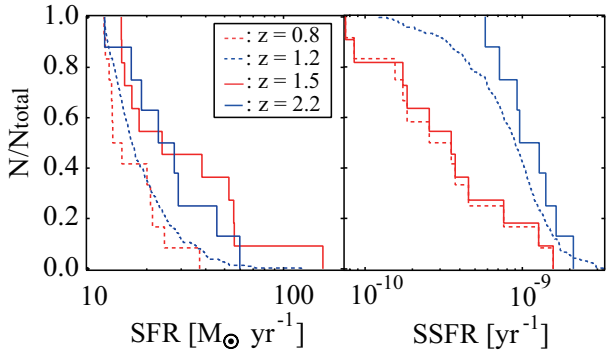}
\caption{The cumulative functions of SFR and SSFR. Red and blue lines show cluster and field samples, respectively: the H$\alpha $ emitters at $z=0.8$ in RXJ1716 (red dashed line; \citealt{2010MNRAS.403.1611K}), [OII] emitters at $z=1.2$ in COSMOS field (blue solid line; \citealt{2009ApJ...700..971I}), [OII] emitters at $z=1.5$ in XCS2215 (red solid line; \citealt{2010MNRAS.402.1980H}) and H$\alpha $ emitters at $z=2.2$ in GOODS-N field (blue dashed line; This work).}
\label{fig9}
\end{center}
\end{figure}

The results presented in this paper on the time evolution and the environmental dependence of star forming activities are all intriguing. However, we are still very limited by the small area and/or the small number of observed fields and clusters. We are conducting the MAHALO-Subaru project (MApping HAlpha and Lines of Oxygen with Subaru; Kodama et al.\ in preparation) which aims to map out star forming activities over a wide redshift range ($0.4<z<2.5$) and in various environments by targeting many distant clusters and some general fields. By using a unique, large set of narrow-band filters available on the wide field cameras, Suprime-Cam and MOIRCS on Subaru Telescope, we can build a statistical, unbiased sample of line emitting galaxies (H$\alpha $ and [OII]). This massive data set will enable us to sketch how the star formation activities propagate in the universe over the cosmic time and across environments, with much greater statistics and smaller uncertainties, hence providing more comprehensive views of galaxy evolution in their most active phase of star formation and mass assembly.

\begin{figure}
\begin{center}
  \FigureFile(80mm,50mm){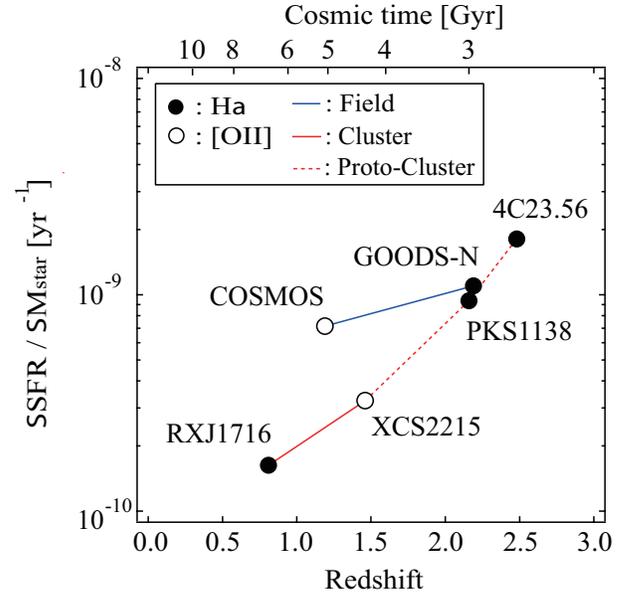}
\caption{The evolution of star formation activities of the line emitting galaxies probed by narrow-band surveys. Filled circles and open circles indicate H$\alpha $ emitters and [OII] emitters, respectively. The 4C23.56 sample includes only the galaxies with large SFRs ($>25$M$_\odot $yr$^{-1}$).}
\label{fig10}
\end{center}
\end{figure}

\section{Summary}

We have conducted a narrow-band imaging and spectroscopic surveys of H$\alpha $ emitters at $z=2.2$ in GOODS-North field, using MOIRCS on Subaru. Our survey has identified 11 H$\alpha $ emitters and one AGN over a 70 arcmin$^2$ area.
We have confirmed probable H$\alpha $ emission lines for all of the seven targets by the spectroscopic follow-up observation, on top of the two already confirmed star forming galaxies at $z_{\mathrm{spec}}=2.2$.  Therefore, our technique of searching for H$\alpha $ emitters at $z=2.2$ based on the excess fluxes in the narrow-band (NB209) and photometric redshifts, is proven to be robust and efficient. The results and conclusions of this study are summarized below:

\begin{enumerate}

\item The H$\alpha $ emitters have SFRs ranging from 12~M$_\odot \mathrm{yr}^{-1}$ to 60~M$_\odot \mathrm{yr}^{-1}$, with the mean SFR of $\langle SFR \rangle =27.8$M$_\odot \mathrm{yr}^{-1}$.  Note that we have corrected for dust extinction by assuming the typical value of A(H$\alpha $)=1.
The averaged stellar mass of the H$\alpha $ emitters is $4.0\times 10^{10}$~M$_\odot$, and we find the correlation between the stellar mass and the specific star formation rate
in the sense that more massive galaxies tend to have lower specific star formation rates.

\item The H$\alpha $ luminosity function is derived from our data by combining the data points reproduced from the previous works in the literature for the brighter magnitudes. The combined LF is represented by a Schechter function with log$L$ = 42.82, log$\phi = -2.78$ and $\alpha = -1.38$. Our result shows a moderate steepness of the faint-end slope. By integrating the luminosity function thus derived, we find that the cosmic star formation rate at $z=2.2$ is $\rho _{\mathrm{SFR}}=0.31$~M$_\odot$yr$^{-1}$Mpc$^{-3}$, which is consistent with other previous studies at $z\sim 2$.

\item We have compared the properties of our emitters in the general field or low density environment, with those in the cluster environments to investigate the environmental dependence of galaxy evolution. There is a difference in the degree of time evolution of $\Sigma$SFR / $\Sigma M_{\mathrm{star}}$ between the two environments. This implies that the star formation activity is enhanced at $z>2$ in high density regions as a consequence of ``galaxy formation bias'' in the early universe.

\end{enumerate}

We must warn however that there is a possibility that these results are still severely affected by the cosmic variance, which may well be expected from the actual inhomogeneous spatial distribution of our H$\alpha $ emitters. It is necessary to extend the survey area to cover a representative volume of the universe and average over the cosmic variance in order to obtain more robust conclusions.

\section*{Acknowledgments}
This paper is based on data collected at Subaru Telescope, which is operated by the National Astronomical Observatory of Japan. We thank the Subaru telescope staff for their help in the observation. We are also very grateful to the MODS team for allowing us to use MODS data and catalog and Yuko Ideue for providing information about [OII] emitters at $z=1.2$ in COSMOS field.
We thank the anonymous referee who gave us many useful comments, which improved the paper.
T.K. acknowledges the financial support in part by a Grant-in-Aid for the Scientific Research (Nos.\, 18684004 and 21340045) by the Japanese Ministry of Education, Culture, Sports, Science and Technology.
Y.K. acknowledge the support from the Japan Society for the Promotion of Science (JSPS) through JSPS research fellowships for young scientists.  
\bibliographystyle{mn2e}
\bibliography{ref}

\end{document}